\documentclass[pre,twocolumn,showpacs]{revtex4}
\newcommand{\g}[1]{\mbox{\boldmath ${#1}$}}

\begin{document}

\title{Variational principle for frozen-in vortex structures
interacting with sound waves}
\author{V. P. Ruban}
\email{ruban@itp.ac.ru}
\affiliation{$^{1}$L.D.Landau Institute for Theoretical Physics,
2 Kosygin Street, 119334 Moscow, Russia}
\date{\today}

\begin{abstract}
General properties of conservative hydrodynamic-type models are treated
from positions of the canonical formalism adopted for liquid continuous media,
with applications to the compressible Eulerian hydrodynamics,
special- and general-relativistic fluid dynamics,
and two-fluid plasma model including the Hall-magnetohydrodynamics.
A variational formulation is found for motion and interaction of
frozen-in localized vortex structures and acoustic waves in a special
description where dynamical variables are, besides the Eulerian fields
of the fluid density and the potential component of the canonical momentum,
also the shapes of frozen-in lines of the generalized vorticity.
This variational principle can serve as a basis for approximate dynamical
models with reduced number of degrees of freedom.

\end{abstract}

% 47.10.+g General theory
% 47.15.Ki Inviscid flows with vorticity
% 47.32.Cc Vortex dynamics
% 47.75.+f Relativistic fluid dynamics
% 52.30.Ex Two-fluid and multi-fluid plasmas
% 52.30.Cv Magnetohydrodynamics (including electron magnetohydrodynamics)
% 52.30.-q Plasma dynamics and flow
% 04.20.Fy Canonical formalism, Lagrangians, and variational principles
% 04. General relativity and gravitation

\maketitle

\section{Introduction}

In many physical systems the motion of continuous liquid media can be
approximately described by hydrodynamic-type equations
having a remarkable mathematical structure based on an underlying
variational least action principle in the Lagrangian description
\cite{LL2, ZK97, R99, R2000PRD, R2001PRE, RP2001PRD, R2002}.
The characteristic
feature of the hydrodynamic-type systems is that they possess, besides the
usual integrals of motion (the total energy, the total linear momentum,
and the total angular momentum), also an infinite number of specific
integrals of motion related to the freezing-in property of canonical
vorticity \cite{ZK97,R99,R2001PRE}.
%The reason for this is a basic
%physical property of fluids, relabeling symmetry.
Thus, hydrodynamic equations  describe an interaction
between "soft" degrees of freedom of  -- frozen-in vortices,
and "hard" degrees of freedom -- acoustic modes. However, in the Eulerian
description of flows by the density and by the velocity (or the canonical
momentum) fields,
vortices and sound waves are "mixed". Another point is that due to unresolved
freezing-in constraints, the Eulerian equations of motion do not follow
directly from a variational principle (see \cite{ZK97} for discussion).
This work has two main purposes. The first purpose is to introduce a such
general description of ideal flows, that soft and
hard degrees of freedom are explicitly separated, and the frozen-in property 
of the vorticity is taken into account. The second purpose is
to formulate a principle of least action in this representation.
As the result, the acoustic waves will be described by the Eulerian fields
of the fluid density and the potential component of the canonical momentum,
while the canonical vorticity will be represented as a continuous distribution 
of frozen-in vortex lines (the so called formalism of vortex lines
\cite{R99,R2000PRD,R2001PRE}, that previously was applied only to static 
density profiles). The Lagrangian of this dynamical
system is a non-trivial unification of the canonical Lagrangian
corresponding to purely potential flows, with a generalized Lagrangian of
vortex lines.

\section{Canonical formalism for fluids}

\subsection{Generalized Euler equation}

Typically in a complex classical system on the microscopic
level there are permanently existing particles of several kinds,
for instance, molecules in a gas, or the electrons and ions in a plasma.
In general situation, different components can have different macroscopically
averaged velocities near a same point and/or different relative
concentrations in separated points.
In such cases, each population of the complex fluid should be included into
consideration individually, for example, as in the two-fluid plasma model 
discussed some later in this work. Now for simplicity we are going to 
consider the case when the macroscopic velocities of all components coincide
and mutual relations between the concentrations are homogeneous
in space and time, so the macroscopically averaged physical state of
the medium at a given point $\g{r}=(x,y,z)$ at a given time moment $t$ is
completely determined by two quantities, namely by a scalar $n(\g{r},t)$,
which is proportional to concentration of conservative particles
of a definite sort, and by a vector $\g{j}(\g{r},t)$, the corresponding
density of flow. The field  $\g{j}$ is related by the continuity equation 
to the field $n$,
\begin{equation}\label{continuity}
n_t+\mbox{div\,}\g{j}=0,
\end{equation}
where subscript is used to denote the  partial derivative.
It is clear that $\g{j}=n\g{v}$, where $\g{v}(\g{r},t)$ is the macroscopic
velocity field.
Let each point of the fluid medium be marked by a label $\g{a}=(a_1,a_2,a_3)$,
so the mapping $\g{r}=\g{x}(\g{a},t)$ is the full Lagrangian description of 
the flow. The less exhaustive description of the flow by the fields $n(\g{r},t)$ 
and $\g{j}(\g{r},t)$ is commonly referred as the Eulerian description.
The relations between the Eulerian fields and the Lagrangian mapping
are the following,
\begin{eqnarray}\label{n_x}
&&n(\g{r},t)=\int\delta(\g{r}-\g{x}(\g{a},t))d\g{a},\\
\label{j_x}
&&\g{j}(\g{r},t)=\int\delta(\g{r}-\g{x}(\g{a},t))\g{x}_t(\g{a},t)d\g{a},
\end{eqnarray}
and they satisfy the continuity equation (\ref{continuity}) automatically.

With neglecting all dissipative processes (due to viscosity, diffusion, etc.), 
and assuming internal properties of the fluid homogeneous (such as the 
specific entropy in adiabatic flows, or the temperature in isothermal flows), 
the trajectories $\g{r}=\g{x}(\g{a},t)$ of fluid elements are determined by the 
variational principle $\delta(\int \tilde{\cal L} dt)/\delta \g{x}(\g{a},t)=0$,
with the Lagrangian of a special general form, actually depending only
on the Eulerian fields $n(\g{r},t)$ and $\g{j}(\g{r},t)$,
\begin{equation}\label{L_deneral_form}
\tilde{\cal L}\{\g{x}(\g{a}),\g{x}_t(\g{a})\}=
{\cal L}\{n(\g{r}), \g{j}(\g{r})\}|_{n\{{\bf x}\},j\{{\bf x},{\bf x}_t\}},
\end{equation}
where the braces $\{\dots\}$ are used to denote functional arguments
as against usual scalar or vector arguments that are denoted by the 
parenthesis $(\dots)$.
The equation of motion, corresponding to the Lagrangian (\ref{L_deneral_form}),
has a remarkable general structure. The usual variational Euler-Lagrange
equation,
$$
\frac{d}{dt}\frac{\delta \tilde{\cal L}}{\delta \g{x}_t(\g{a})}=
\frac{\delta \tilde{\cal L}}{\delta \g{x}(\g{a})},
$$
in the Eulerian representation has the form (generalized Euler equation)
\begin{equation}\label{Euler_general_compressible}
\frac{\partial}{\partial t}\left(\frac{\delta {\cal L}}{\delta\g{j}}\right)=
\left[\frac{\g{j}}{n}\times
\mbox{curl}\left(\frac{\delta {\cal L}}{\delta\g{j}}\right)
\right]+\g{\nabla}\left(\frac{\delta {\cal L}}{\delta n}\right),
\end{equation}
where the variational derivative ${\delta {\cal L}}/{\delta\g{j}}$ is taken
at fixed $n(\g{r},t)$, while the variational derivative
${\delta {\cal L}}/{\delta n}$ is taken at fixed $\g{j}(\g{r},t)$
(compare with Refs. \cite{R99,R2001PRE, R2000PRD}, where this equation is 
written in terms of $n$ and $\g{v}$ and thus looks differently).
Eq.(\ref{Euler_general_compressible}) together with the continuity equation
(\ref{continuity}) completely determine the time evolution of the fields
$n(\g{r},t)$ and $\g{j}(\g{r},t)$ .

\subsection{Hamiltonian structure}

In the Hamiltonian description adopted for fluids as it is discussed in
\cite{R99,R2001PRE}, instead of the field $\g{j}$
the variational derivative of the Lagrangian,
\begin{equation}\label{p_definition}
\g{p}= \frac{\delta {\cal L}}{\delta\g{j}},
\end{equation}
is used (the canonical momentum). The Hamiltonian functional
is defined as the Legendre transformation,
\begin{equation}\label{H_definition}
{\cal H}\{n,\g{p}\}\equiv\int\left(
\frac{\delta {\cal L}}{\delta\g{j}}\cdot\g{j}\right)d\g{r}-{\cal L},
\end{equation}
where $\g{j}$  should be expressed in terms of $\g{p}$ and $n$.
The equations of motion (\ref{Euler_general_compressible}) and 
(\ref{continuity}) now have the non-canonical Hamiltonian structure
\cite{ZK97,R99,R2001PRE},
\begin{eqnarray}
&&\g{p}_t=
\left[\frac{1}{n}\Big( \frac{\delta {\cal H}}{\delta\g{p}}\Big) \times
\mbox{curl\,}\g{p}
\right]-\g{\nabla}\left(\frac{\delta {\cal H}}{\delta n}\right),
\label{p_t_Ham_noncanon_compress}\\
&&n_t=-\mbox{div}\Big( \frac{\delta {\cal H}}{\delta\g{p}}\Big).
\label{n_t_Ham_noncanon_compress}
\end{eqnarray}
These equations can be written as
$\g{p}_t=\{\g{p}, {\cal H} \}$ and $n_t=\{n, {\cal H} \}$, where
the non-canonical Poisson bracket is given by the following expression
(see \cite{ZK97, R99} and references therein about details),
\begin{eqnarray}
&&\{{\cal F}, {\cal H} \}=
\int\left[\frac{\delta {\cal H}}{\delta n}
\left(\g{\nabla}\cdot\frac{\delta {\cal F}}{\delta \g{p}}\right)
-\frac{\delta {\cal F}}{\delta n}
\left(\g{\nabla}\cdot\frac{\delta {\cal H}}{\delta \g{p}}
\right)\right]d\g{r}\nonumber\\
&&\qquad\qquad+\int\left(\frac{\mbox{curl\,}\g{p}}{n}\cdot\left[
\frac{\delta {\cal F}}{\delta \g{p}}\times
\frac{\delta {\cal H}}{\delta \g{p}}\right]\right)d\g{r}.
\label{Poisson_noncanon}
\end{eqnarray}

\section{Particular examples}

To emphasize universality of the employed approach, now we are going to 
consider several physically interesting examples.

\subsection{Eulerian hydrodynamics}

Let us start with the usual Eulerian hydrodynamics.
In this simple case $n$ is the density of the fluid, and the Lagrangian is
the difference between the total macroscopic kinetic energy and 
the total potential energy including the thermal internal energy,
$$
{\cal L}_E=\int \Big(\frac{\g{j}^2}{2n}-\varepsilon(n)
-n U(\g{r},t)\Big)d\g{r},
$$
where $\varepsilon(n)$ is the density of the internal energy, and
$ U(\g{r},t)$ is the potential of an external force.

The canonical momentum coincides with the velocity field,
$$
\g{p}=\frac{\g{j}}{n}=\g{v},
$$
and the Hamiltonian is the total energy expressed in terms of $n$ and $\g{p}$,
$$
{\cal H}_E=\int \Big(n\frac{\g{p}^2}{2}+\varepsilon(n)
+n U(\g{r},t)\Big)d\g{r}.
$$
The equations of motion 
(\ref{p_t_Ham_noncanon_compress}-\ref{n_t_Ham_noncanon_compress})
with this Hamiltonian take the known form
\begin{eqnarray}\label{dp_dt_Euler}
&&\g{p}_t=\left[\g{p}\times \mbox{curl\,}\g{p}\right]
-\g{\nabla}\Big( \frac{\g{p}^2}{2}+ \varepsilon'(n)+U(\g{r},t)\Big),
\nonumber\\
&&n_t=-\mbox{div}( n \g{p}).\nonumber
\label{d_n_dt_Euler}
\end{eqnarray}

\subsection{Relativistic fluid dynamics}

In the Special Relativity  (with $c=1$ for simplicity)
the field $n$ is the time component of the (contra-variant)
4-vector of current, while $\g{j}$ is the space component \cite{LL6}.
The absolute value $\tilde n$ of this 4-vector is equal to 
$(n^2-\g{j}^2)^{1/2}$ and it has the meaning of the concentration of
conservative particles in the locally co-moving frame of reference.
The invariant expression for the action functional implies the Lagrangian 
in the form (compare with \cite{R99})
$$
{\cal L}_r=-\int\varepsilon\left((n^2-\g{j}^2)^{1/2}\right)d\g{r},
$$
where $\varepsilon(\tilde n)$ is the relativistic density of the internal
fluid energy including the rest energy. The canonical momentum field is 
defined by the relation
$$
\g{p}=\varepsilon'\Big((n^2-\g{j}^2)^{1/2}\Big)
\frac{\g{j}}{(n^2-\g{j}^2)^{1/2}}.
$$
Unlike the Eulerian hydrodynamics, now it is not possible in general to
get analytically the inverse relation $\g{j}(n,\g{p})$ in order to
substitute it into the expression for the Hamiltonian density,
$$
 h_r=\varepsilon'\Big((n^2-\g{j}^2)^{1/2}\Big)
\frac{\g{j}^2}{(n^2-\g{j}^2)^{1/2}}+
\varepsilon\Big((n^2-\g{j}^2)^{1/2}\Big).
$$
Exceptions are some special dependences $\varepsilon(\tilde n)$
(see, for example, \cite{RP2001PRD} where the particular case
$\varepsilon(\tilde n)\propto \tilde n^{4/3}$ is considered,
corresponding to the ultra-relativistic equation of state).

In the General Relativity the continuity equation is (see \cite{LL2,LL6})
$$
\frac{1}{\sqrt{-g}}\frac{\partial}{\partial x^i}\left(
\sqrt{-g}\tilde n \frac{dx^i}{ds}\right)=0,
$$
where ${dx^i}/{ds}$ is the 4-velocity of the fluid element passing through 
the point $(t,\g{r})$,
and $g=\mbox{det}\|g_{ik}\|$, the determinant of the metric
tensor $g_{ik}(t,\g{r})$. Therefore
$$
n=\sqrt{-g}\tilde n \frac{dt}{ds},\qquad
j^\alpha=\sqrt{-g}\tilde n \frac{dx^\alpha}{ds},
$$
\begin{equation}
\tilde n =\sqrt{(g_{00}n^2+2g_{0\alpha}nj^\alpha+
g_{\alpha\beta}j^\alpha j^\beta)}/\sqrt{-g},
\end{equation}
and the Lagrangian of the general-relativistic hydrodynamics is
\begin{equation}
{\cal L}_{g.r.} =-\int\varepsilon\left(
\frac{\sqrt{g_{00}n^2+2g_{0\alpha}nj^\alpha+
g_{\alpha\beta}j^\alpha j^\beta}}{\sqrt{-g}}\right)\sqrt{-g}d\g{r}.
\end{equation}
The canonical momentum
$$
p_\alpha=\varepsilon'(\tilde n)\frac{-(g_{0\alpha}n+g_{\alpha\beta}j^\beta)}
{\sqrt{g_{00}n^2+2g_{0\alpha}nj^\alpha+
g_{\alpha\beta}j^\alpha j^\beta}}
$$
depends in a complicated manner on $n$, $\g{j}$, $g_{ik}$.
This circumstance makes impossible in general case
to present an analytical expression for the
corresponding Hamiltonian functional, but, of course, it cannot cancel
the existence of the Hamiltonian in mathematical sense.

\subsection{Two-fluid plasma model}

Analogously multi-component hydrodynamical models can be investigated
where several fields $n^a$ and $\g{j}^a$ are present corresponding
to different sorts of particles, with $a =1,2,\dots, A$.
The Hamiltonian non-canonical equations of motion for such models
have the same general structure as discussed above, and they should
be written for each component. Below we consider a physically important
example --- the two-fluid plasma model. As special limit cases, this model
contains the usual magnetohydrodynamics (MHD), the Electron MHD, 
and the Hall MHD.

\subsubsection{Lagrangian formalism}

As the start point in investigation the two-fluid plasma model,
let us consider the microscopic Lagrangian of a system of electrically 
charged classical point particles, as it is given in the famous book by 
Landau and Lifshitz \cite{LL2}. This Lagrangian is approximately valid up to
the second order on $v/c$ since excitation of the free electro-magnetic field 
by moving charges is negligible,
\begin{eqnarray}
&&{\cal L}_{\mbox{\scriptsize micro}}=\sum_a\frac{m_a\g{v}_a^2}{2}
-\frac{1}{2}\sum_{a\not =b}\frac{e_a e_b}{|\g{r}_a-\g{r}_b|}
+\sum_a\frac{m_a\g{v}_a^4}{8c^2}\nonumber\\
&&+\frac{1}{4c^2}\!\sum_{a\not =b}\frac{e_a e_b}{|\g{r}_a\!-\!\g{r}_b|}
\Big(\g{v}_a\cdot\g{v}_b+
(\g{v}_a\cdot\g{n}_{ab})(\g{v}_b\cdot\g{n}_{ab})\Big),
\label{Lmicro}
\end{eqnarray}
where $\g{r}_a(t)$ are the positions of the point charges $e_a$,
$\g{v}_a(t)\equiv\dot{\g{r}}_a(t)$ are their velocities,
$\g{n}_{ab}(t)$ are the unit vectors in the direction between $e_a$ and $e_b$,
$$
\g{n}_{ab}=\frac{\g{r}_a-\g{r}_b}{|\g{r}_a-\g{r}_b|}.
$$
The first double sum in Eq.(\ref{Lmicro}) corresponds to the electrostatic
interaction, while the second double sum
describes the magnetic interaction via quasi-stationary magnetic field.
It is very important that for a system with macroscopically huge number
of particles the magnetic energy can be of the same order (or even larger)
as the macroscopic kinetic energy
produced by the first ordinary sum in Eq.(\ref{Lmicro}),
while the terms of the fourth order on the velocities are often negligible.
Generally speaking, a large part of plasma physics is governed by this
Lagrangian, at least in the cases when the velocities of particles are
non-relativistic
and the free electro-magnetic field is not excited significantly. Obviously,
different physical problems need different procedures of macroscopic
consideration of this system. The most accurate (and the most complicated)
would be a kinetic description. However, for our purposes it is sufficient
to apply more simple and naive procedure of the hydrodynamical averaging,
that gives less accurate description of the system in terms of the
concentration $n(\g{r},t)$ of electrons  and the density $\g{j}(\g{r},t)$
of their flow, that satisfy the continuity equation,
$n_t+\mbox{div\,}\g{j}=0$ (and corresponding fields
$N(\g{r},t)$, $\g{J}(\g{r},t)$ for the ions, normalized to one elementary
electric charge $e$, so $N=ZN_i$.)

Neglecting all dissipative processes that take place due to collisions
of the particles (though on this step we strongly reduce applicability of the
following conservative two-fluid plasma model), we derive from
Eq.(\ref{Lmicro}) the following Lagrangian functional
\begin{eqnarray}  \label{LagrHallMHD}
&&{\cal L}_{2f}=\int\Big[\frac{M}{2N}\g{J}^2  +\frac{m}{2n}\g{j}^2
+\frac{2\pi e^2}{c^2}
[\mbox{curl}^{-1}(\g{J}-\g{j})_\perp]^2\Big]d\g{r} \nonumber\\
&&-\frac{e^2}{2}\!\int\!\!
\int \!\frac{d\g{r}_1d\g{r}_2}{|\g{r}_1-\g{r}_2|}
\Big(N(\g{r}_1)-n(\g{r}_1)\Big)\Big(N(\g{r}_2)-n(\g{r}_2)\Big)\nonumber\\
&&-\int \left[ T_e n\ln \frac n{f(T_e)}
+T_i\frac{N}{Z}\ln \frac N{ZF(T_i)} \right]d\g{r} ,
\end{eqnarray}
where $(\g{J}-\g{j})_\perp$ is the divergence-free component of the
total current. Here the constant $M$ is the ion mass per one elementary
electric charge, $M=M_i/Z$. The electron mass is $m\ll M$, and it will
be neglected where possible.  The magnetic energy
$\int (\g{B}^2/8\pi)d\g{r}$ is included into this Lagrangian, where
the magnetic field is
\begin{equation}\label{magnetic field}
\g{B}=\frac{4\pi e}{c}\mbox{curl}^{-1}(\g{J}-\g{j})_\perp.
\end{equation}
The terms with $T_e n\ln n$ and $(T_i/Z) N\ln N$
(approximate expressions for the densities of the thermal free energy,
\cite{LL5})
have been introduced in order the macroscopic equations of motion
to contain the pressure terms like $-\nabla p/n$
(see the last term in Eq.(\ref{Euler_general_compressible})),
where $p\approx nT_e$ is the pressure of the hot electron gas,
which is supposed to be isothermal with a temperature $T_e$.
The functions $f(T_e)$ and $F(T_i)$ actually are not important
since they do not contribute to the equations of motion
and thus will be omitted in further equations.

It should be kept in mind that densities of the internal energy
should be used instead of densities of the free energy if
we suppose the flows to be isentropic.
However, since the thermal conductivity is large at high temperatures
($\kappa\propto T_e^{5/2}/e^4m^{1/2}$, \cite{LL10}),
the isothermal approximation usually works better than isentropic one.

\subsubsection{Hamiltonian formalism. Hall MHD limit}

For two-fluid models the Hamiltonian functional is defined as follows,
\begin{equation}\label{H_definition_two_fluid}
{\cal H}_{2f}\{n,\g{p},N,\g{P}\}\equiv\int\left(
\frac{\delta {\cal L}_{2f}}{\delta\g{j}}\cdot\g{j}+
\frac{\delta {\cal L}_{2f}}{\delta\g{J}}\cdot\g{J}
 \right)d\g{r}-{\cal L}_{2f},
\end{equation}
where $\g{j}$ and $\g{J}$ should be expressed in terms of the electron and ion
canonical momenta $\g{p}$ and $\g{P}$.

In our particular case  we have
\begin{eqnarray}
\g{P}&=&\frac{M\g{J}}{N}+
\frac{4\pi e^2}{c^2}\mbox{curl}^{-2}(\g{J}-\g{j})_\perp,\label{P_HallMHD} \\
\g{p}&=&\frac{m\g{j}}{n}-
\frac{4\pi e^2}{c^2}\mbox{curl}^{-2}(\g{J}-\g{j})_\perp,\label{p_HallMHD}
\end{eqnarray}
and this results in the following approximate (valid in the limit of small
$m$) expression for the Hamiltonian,
\begin{eqnarray}  \label{HamiltonianHallMHD}
&&\!\!{\cal H}_{2f}\approx\!\int\!\!\Big[\frac{N}{2M}(\g{P}+\g{p})^2 +
\frac{n}{2m}\g{p}_\parallel^2  +
\frac{c^2}{8\pi e^2}(\mbox{curl\,}\g{p})^2\Big]d\g{r} \nonumber\\
&&+\frac{e^2}{2}\!\int\!\!
\int \!\frac{d\g{r}_1d\g{r}_2}{|\g{r}_1-\g{r}_2|}
\Big(N(\g{r}_1)-n(\g{r}_1)\Big)\Big(N(\g{r}_2)-n(\g{r}_2)\Big)\nonumber\\
&&+\int [T_e n\ln n +(T_i/Z) N\ln N]d\g{r} ,
\end{eqnarray}
where $\g{p}_\parallel$ is the potential component of the electron
canonical momentum field.

The equations of motion for the electron component
 now have the non-canonical Hamiltonian structure
(\ref{p_t_Ham_noncanon_compress}-\ref{n_t_Ham_noncanon_compress})
(and analogous equations for the ion component).
Due to the large coefficient $\propto 1/m$ in front of $\g{p}_\parallel^2$
in the Hamiltonian,  $\g{p}_\parallel$ should be small.
A small value of $m$ is also the reason for the electrical quasi-neutrality
on sufficiently large scales. Therefore we can put $n\approx N$.
Correspondingly, these simplifications will result in
${\cal H}_{2f}\to {\cal H}_{simpl}\{N,\g{P} +\g{p},\mbox{curl\,}\g{p}\}$.
Thus, we have to deal with $N$, with the ion velocity
$$\g{u}=(\g{P} +\g{p})/M,$$ and with
$$\g{\omega}=\mbox{curl\,}\g{p}/M=-(e/Mc)\g{B}.$$

Let us introduce the dimensionless density
$$\rho(\g{r},t)=N(\g{r},t)/N_0,$$
the ion inertial length $d_i$ at the static homogeneous  state,
$$d_i^2={Mc^2}/({4\pi e^2N_0})=M_ic^2/(4\pi Z^2 e^2 N_{i0}),$$
and the speed $s$ of the ion-acoustic waves,
$$s^2=(T_i+ZT_e)/M_i.$$
The Hamiltonian ${\cal H}_H\{\rho,\g{u},\g{\omega}\}=(1/M){\cal H}_{simpl}$
for this simplified two-fluid plasma model (known as the Hall MHD) is
\begin{equation}\label{H_H}
{\cal H}_H=\int\Big[\frac{\rho\g{u}^2}{2}
+d_i^2\frac{\g{\omega}^2}{2} +s^2 \rho\ln \rho\Big]d\g{r}.
\end{equation}
The equations of motion are
\begin{eqnarray}\label{du_dt_general}
&&\partial_t\g{u}=
\left[\frac{1}{\rho} \Big(\frac{\delta {\cal H}_H}{\delta\g{u}}\Big) \times
\mbox{curl\,}\g{u}\right] \nonumber\\
&&\qquad+ \frac{1}{\rho}\left[
\mbox{curl\,}\Big(\frac{\delta {\cal H}_H}{\delta\g{\omega}}\Big) \times
\g{\omega}
\right]
-\g{\nabla}\left(\frac{\delta {\cal H}_H}{\delta \rho}\right),\\
&&\partial_t \rho=-\mbox{div}\Big( \frac{\delta {\cal H}_H}{\delta\g{u}}\Big),
\label{d_rho_dt_general} \\
&&\partial_t\g{\omega}=
\mbox{curl\,}\left[\frac{1}{\rho}
\Big(\frac{\delta {\cal H}_H}{\delta\g{u}}+
\mbox{curl\,}\frac{\delta {\cal H}_H}{\delta\g{\omega}}\Big)\times\g{\omega}
\right] .
\end{eqnarray}
In explicit form we easily obtain  the Hall MHD equations
as they are commonly known,
\begin{equation}\label{du_dt_HMHD}
\g{u}_t=
\left[\g{u} \times\mbox{curl\,}\g{u}\right]
+\frac{d_i^2}{\rho}\left[\mbox{curl\,}\g{\omega} \times\g{\omega}\right]
-\g{\nabla} \frac{\g{u}^2}{2}-s^2\frac{\g{\nabla} \rho}{\rho},
\end{equation}
\begin{equation}
\rho_t=-\mbox{div}( \rho \g{u}),
\label{d_rho_dt_HMHD}
\end{equation}
\begin{equation}
\g{\omega}_t= \mbox{curl}
\left[\Big(\g{u}+\frac{d_i^2}{\rho}\mbox{curl\,}\g{\omega}\Big) \times
\g{\omega}\right].
\end{equation}
Obviously, one can normalize the length and the velocity scales
to $d_i$ and $s$ respectively, that will result effectively in $d_i=1$,
$s=1$.

It is interesting to note that the ion generalized  vorticity can
be identically equal to zero (that means
$\mbox{curl\,}\g{P}=0$, and therefore $\g{\omega}=\mbox{curl\,}\g{u}$),
and then we have
\begin{eqnarray}\label{du_dt_single}
\!\!&\!\!&\!\!\g{u}_t=
\left[\Big(\g{u}+\frac{\mbox{curl\,}\mbox{curl\,}\g{u}}{\rho}\Big)
\times \mbox{curl\,}\g{u}\right]
-\g{\nabla} \frac{\g{u}^2}{2}-\frac{\g{\nabla} \rho}{\rho},\\
\!\!&\!\!&\!\!\rho_t=-\mbox{div}( \rho \g{u}).
\label{d_rho_dt_single}
\end{eqnarray}
This reduced system formally corresponds to a single-fluid model with the
Hamiltonian ${\cal H}_s\{\rho,\g{u}\}$  as follows,
\begin{equation}
{\cal H}_s=\int\Big[\frac{\rho\g{u}^2}{2}
+\frac{(\mbox{curl\,}\g{u})^2}{2} + \rho\ln \rho\Big]d\g{r},
\end{equation}
because the equations (\ref{du_dt_single}-\ref{d_rho_dt_single})
have the standard form  similar to Eqs.
(\ref{p_t_Ham_noncanon_compress}-\ref{n_t_Ham_noncanon_compress}),
\begin{eqnarray}\label{du_dt_s}
&&\partial_t\g{u}=
\left[\frac{1}{\rho} \Big(\frac{\delta {\cal H}_s}{\delta\g{u}}\Big) \times
\mbox{curl\,}\g{u}
\right]-\g{\nabla}\left(\frac{\delta {\cal H}_s}{\delta \rho}\right),\\
&&\partial_t \rho=-\mbox{div}\Big( \frac{\delta {\cal H}_s}{\delta\g{u}}\Big).
\label{d_rho_dt_s}
\end{eqnarray}

\section{Interaction between frozen-in vortex lines and acoustic modes.}

The Hamiltonian non-canonical equations
(\ref{p_t_Ham_noncanon_compress}-\ref{n_t_Ham_noncanon_compress})
do not follow directly from a variational principle.
The mathematical reason for this is a degeneracy of the corresponding
non-canonical Poisson bracket (\ref{Poisson_noncanon}) which is discussed,
for instance, in \cite{ZK97}.
The degeneracy results in frozen-in property for
the canonical vorticity field $\g{\omega}=\mbox{curl\,}\g{p}$.
However, representations of the canonical momentum in terms of auxiliary 
variables exist that fix topological structure of  vortex lines,
and then a variational formulation becomes possible. A known example
of such auxiliary variables are  the Clebsch variables \cite{ZK97, SBR2003},
when $\g{p}= \g{\nabla}\varphi+(\lambda/n)\g{\nabla}\mu$
and $(n,\varphi)$, $(\lambda,\mu)$ are two pairs of canonically conjugate
variables. But the Clebsch representation usually
is not suitable for studying localized vortex structures like vortex filaments.
Below we consider another representation for the canonical
momentum field, when dynamical variables are the shapes of vortex lines.
For a nearly static density profile, $n\approx n_0(\g{r})$, such description
was used in \cite{R2001PRE, R2000PRD} to study slow flows in spatially 
inhomogeneous systems. Now we are going to introduce a variational formulation
valid for the general case, since the function $n(\g{r},t)$ 
is also an unknown variable. It will be demonstrated that
variational principle with the Lagrangian
(\ref{L_vort_sound}) determines equations of motion for shapes of 
frozen-in vortex lines, for the potential component of the canonical 
momentum field, and for the density profile $n(\g{r},t)$.

So, we decompose the momentum field onto the potential component
and the divergence-free component,
\begin{equation}\label{p_decomposition}
\g{p}(\g{r},t)= \g{\nabla}\varphi(\g{r},t)+\mbox{curl}^{-1}\g{\omega}(\g{r},t).
\end{equation}
Accordingly, the field $\g{j}$ is decomposed,
\begin{equation}\label{j_decomposition}
\g{j}=\frac{\delta {\cal H}}{\delta \g{p}}=
\frac{\delta {\cal H}}{\delta \g{p}_\parallel} +
\mbox{curl}\frac{\delta {\cal H}}{\delta \g{\omega}}
\equiv\g{j}_\parallel +\g{j}_\perp.
\end{equation}
Obviously, the continuity equation results in the relation
\begin{equation}\label{cont_j_parallel}
\g{\nabla}\Delta^{-1}  n_t=-\g{j}_\parallel.
\end{equation}

For the frozen-in vorticity field we use the so called vortex line
representation. In the simplest form when the lines are closed,
it reads as follows (for details and discussion see \cite{R2001PRE}),
\begin{eqnarray}\label{VLR}
&&\g{\omega}(\g{r},t)=
\int_{\cal N} d^2\nu\oint\delta(\g{r}-\g{R}(\nu,\xi,t))
\g{R}_\xi(\nu,\xi,t)\,d\xi \nonumber\\
&&\qquad\qquad= \frac{\g{R}_\xi(\nu,\xi,t)}
{\mbox{det}\|\partial\g{R}/\partial(\nu,\xi)\|}\Big|_{{\bf R}={\bf r}},
\end{eqnarray}
where the label $\nu=(\nu_1,\nu_2)\in{\cal N}$ belongs to a 2D manifold 
${\cal N}$ and singles out an individual vortex line,
while an arbitrary longitudinal parameter $\xi$ determines a point on the line.
The Jacobian of the mapping $\g{R}(\nu,\xi,t)$ is denoted as
$\mbox{det}\|\partial\g{R}/\partial(\nu,\xi)\|=
([\g{R}_{\nu_1}\times\g{R}_{\nu_2}]\cdot\g{R}_\xi)$.

The divergence-free component of the canonical momentum field now is
given by the expression
\begin{equation}
\g{p}_\perp=\mbox{curl}^{-1}\g{\omega}(\g{r},t)=
\int \frac{[\g{R}_\xi\times(\g{r}-\g{R})]
d^2\nu d\xi}{4\pi|\g{r}-\g{R}|^3}.
\end{equation}

The vorticity variation $\delta\g{\omega}(\g{r},t)$ induced by a variation
$\delta\g{R}(\nu,\xi,t)$ of the vortex lines takes the form \cite{R2001PRE}
\begin{equation}\label{omega_variation}
\delta\g{\omega}(\g{r},t)=\mbox{curl}_{\bf r}
\int_{\cal N} d^2\nu\oint\delta(\g{r}-\g{R}(\nu,\xi,t))
[\delta\g{R}\times\g{R}_\xi ]\,d\xi,
\end{equation}
which follows directly from Eq.(\ref{VLR}). 

It should be noted that in the case of arbitrary topology of the vortex lines,
one has just to replace in the above expressions
$\g{R}(\nu,\xi,t)\to \g{R}(\g{a},t)$, and
$\g{R}_\xi\, d^2\nu\,d\xi\to (\g{\omega}_0(\g{a})\cdot\g{\nabla}_a)
\g{R}(\g{a},t) d\g{a}$, see \cite{R2001PRE}.

Eq.(\ref{omega_variation}) results in the important relations \cite{R2001PRE},
\begin{equation}\label{deltaH_deltaR}
\frac{\delta {\cal H}}{\delta \g{R}}=[\g{R}_\xi\times\g{j}_\perp(\g{R})],
\end{equation}
\begin{equation}\label{VLR_omega_t}
\g{\omega}_t=
\mbox{curl}_{\bf r}\left[\frac{\g{R}_t\times\g{R}_\xi}
{\mbox{det}\|\partial\g{R}/\partial(\nu,\xi)\|}\right]\Big|_{{\bf R}={\bf r}}.
\end{equation}
Therefore the equation of motion for the vorticity,
$$
\g{\omega}_t=\mbox{curl}_{\bf r}[\g{v}\times\g{\omega}], \qquad
\g{v}=\frac{1}{n}\frac{\delta {\cal H}}{\delta \g{p}},
$$
means
\begin{equation}\label{VLR_R_t}
\left[\frac{\g{R}_t\times\g{R}_\xi}
{\mbox{det}\|\partial\g{R}/\partial(\nu,\xi)\|}\right]\Big|_{{\bf R}={\bf r}}=
\left[\frac{\g{j}_\parallel+\g{j}_\perp}{n}\times\g{\omega}\right]
+\nabla_{\bf r}\Psi(\nu),
\end{equation}
where $\Psi(\nu_1,\nu_2)$ is some arbitrary function of two variables.
A possible choice is $\Psi=0$,
but for general purposes we will consider below $\Psi\not =0$.

With using Eqs.(\ref{j_decomposition}), (\ref{cont_j_parallel}),
(\ref{deltaH_deltaR}), (\ref{VLR_R_t}),
one can verify that if the quantities
$\g{R}(\nu,\xi,t)$, $n(\g{r},t)$, and $\varphi(\g{r},t)$ obey equations of motion
corresponding to the following Lagrangian,
\begin {eqnarray}
&&{\cal L}_{\mbox{\scriptsize v-s}}=-\int n \varphi_t d\g{r}
-{\cal H}\{n,\g{\nabla}\varphi+\mbox{curl}^{-1}\g{\omega}\{\g{R}\}\}
\nonumber\\
&&+\int\delta(\g{r}-\g{R}(\nu,\xi,t))
([\g{R}_\xi\times\g{R}_t]\cdot\g{\nabla}_{\bf r}
\Delta^{-1}_{\bf r} n )
d^2\nu\,d\xi\,d\g{r}\nonumber\\
&&-\int\Psi(\nu_1,\nu_2)([\g{R}_{\nu_1}\times\g{R}_{\nu_2}]\cdot\g{R}_\xi)n(\g{R})
d^2\nu\,d\xi,
\label{L_vort_sound}
\end{eqnarray}
then equations (\ref{p_t_Ham_noncanon_compress}-\ref{n_t_Ham_noncanon_compress})
are satisfied. Indeed, the variation of ${\cal L}_{\mbox{\scriptsize v-s}}$ by
$\delta\g{R}(\nu,\xi,t)$ gives the equation
\begin{eqnarray}
&&[\g{R}_\xi\times\g{R}_t]\,n(\g{R})-[\g{R}_\xi\times\g{j}_\parallel(\g{R})]\nonumber\\
&&\quad=\frac{\delta {\cal H}}{\delta \g{R}}-
n(\g{R})\mbox{det}\Big\|\frac{\partial\g{R}}{\partial(\nu,\xi)}\Big\|
\nabla_{\bf r}\Psi(\nu),
\end{eqnarray}
which is easily recognized as Eq.(\ref{VLR_R_t}),
\begin{eqnarray}
&&[\g{R}_t\times\g{R}_\xi]=
\left[\frac{\g{j}_\parallel(\g{R}) +\g{j}_\perp(\g{R})}
{n(\g{R})}\times\g{R}_\xi\right]
\nonumber\\
&&\qquad\qquad+\mbox{det}\Big\|\frac{\partial\g{R}}{\partial(\nu,\xi)}\Big\|
(\nabla_{\bf r}\Psi)|_{{\bf r}={\bf R}}.
\end{eqnarray}

Variation by $\delta n(\g{r},t)$ results in the potential component of
the Eq.(\ref{p_t_Ham_noncanon_compress}),
$$
-\varphi_t+\Delta^{-1}_{\bf r}\g{\nabla}_{\bf r}\cdot
\left[\frac{\g{R}_t\times\g{R}_\xi}
{\mbox{det}\|\partial\g{R}/\partial(\nu,\xi)\|}\right]\Big|_{{\bf R}={\bf r}}
=\frac{\delta {\cal H}}{\delta n}+\Psi.
$$

Finally, the variation by $\delta \varphi(\g{r},t)$ gives the
continuity equation
$$
n_t+\g{\nabla}\cdot\g{j}_\parallel=0.
$$

Thus, the Lagrangian (\ref{L_vort_sound}) gives a required
variational formulation for the problem of motion and interaction
between localized frozen-in vortex structures (described by the mapping
$\g{R}(\nu,\xi,t)$) and acoustic degrees of freedom (described by the fields
$n(\g{r},t)$ and $\varphi(\g{r},t)$).
This variational principle definitely can serve as a basis for future
approximate analytical and numerical studies dealing with reduced
dynamical systems where only most relevant degrees of freedom will be taken
into account.

The function $\Psi(\nu_1,\nu_2)$ can be useful to investigate 
nearly stationary flows, since the effective Hamiltonian,
$$
\tilde{\cal H}= {\cal H}+
\int\Psi(\nu_1,\nu_2)([\g{R}_{\nu_1}\times\g{R}_{\nu_2}]\cdot\g{R}_\xi)n(\g{R})
d^2\nu\,d\xi,
$$
has an extremum on stationary flows with the velocity field $\g{v}$ 
everywhere directed along vortex surfaces. However, one should remember that
existence of globally defined vortex surfaces (and thus the function
$\Psi$) is an exceptional case in the variety of 3D vector fields.
In the general case one should use $(\g{\omega}_0(\g{a})\cdot\g{\nabla}_a)
\g{R}(\g{a},t) d\g{a}$ instead of $\g{R}_\xi(\nu,\xi,t)d^2\nu d\xi$
in the Lagrangian (\ref{L_vort_sound}) and no function $\Psi(\nu_1,\nu_2)$,
since the labels $\nu$ are not defined.

Generalization of the above theory for multi-fluid models is straightforward.

As an explicit example, below is given the Hamiltonian of the Eulerian
hydrodynamics in terms of $\g{R}(\nu,\xi,t)$, $n(\g{r},t)$, and
$\varphi(\g{r},t)$,
\begin{eqnarray}
&&{\cal H}_E=\int\frac{n}{2}\left(\g{\nabla}\varphi +
\int \frac{[\g{R}_\xi\times(\g{r}-\g{R})]d^2\nu d\xi}
{4\pi|\g{r}-\g{R}|^3}\right)^2d\g{r}\nonumber\\
&&\qquad+\int [\varepsilon(n)+nU(\g{r},t)]d\g{r}.
\end{eqnarray}

\subsection{The case of the Hall MHD}

In the Hall MHD there are two frozen-in generalized vorticity fields but
only one density field. It is easy to check that in this case
the corresponding Lagrangian takes the following form,
\begin {eqnarray}
&&{\cal L}_{\mbox{\scriptsize HMHD}}=-\int \rho \varphi_t d\g{r} \nonumber\\
&&+\int\delta(\g{r}-\g{R}(\nu,\xi,t))
([\g{R}_\xi\times\g{R}_t]\cdot\g{\nabla}_{\bf r}
\Delta^{-1}_{\bf r} \rho ) d^2\nu\,d\xi\,d\g{r}\nonumber\\
&&+\int\delta(\g{r}-\g{X}(\mu,\eta,t))
([\g{X}_\eta\times\g{X}_t]\cdot\g{\nabla}_{\bf r}
\Delta^{-1}_{\bf r} \rho )
d^2\mu\,d\eta\,d\g{r}\nonumber\\
&&-{\cal H}_H\{\rho,\g{\nabla}\varphi+
\mbox{curl}^{-1}\g{\omega}\{\g{R}\}+
\mbox{curl}^{-1}\g{\Omega}\{\g{X}\},
\g{\omega}\{\g{R}\}\}\nonumber\\
&&-\int\Psi(\nu_1,\nu_2)([\g{R}_{\nu_1}\times\g{R}_{\nu_2}]\cdot\g{R}_\xi)
\rho(\g{R})d^2\nu\,d\xi \nonumber\\
&&-\int\Phi(\mu_1,\mu_2)([\g{X}_{\mu_1}\times\g{X}_{\mu_2}]
\cdot\g{R}_\eta)\rho(\g{X})
d^2\mu\,d\eta,
\label{L_v_s_HMHD}
\end{eqnarray}
where the vector function $\g{R}(\nu,\xi,t)$ describes the frozen-in lines
of the electron generalized vorticity $\g{\omega}$,
while $\g{X}(\mu,\eta,t)$ describes the frozen-in lines of the ion 
generalized vorticity $\g{\Omega}$.
The Hamiltonian ${\cal H}_H\{\rho,\g{u},\g{\omega}\}$ is given by
Eq.(\ref{H_H}).

%%%%%%%%%%%%%%%%%%%%%%%%%%%%%%%%%%%%%%%%%%%%%%%%%%%%%%%%%%%%%%%%%%%%%%
\subsection*{Acknowledgments}
These investigations were supported by INTAS (grant No. 00292), by RFBR ,
by the Russian State Program of Support of the Leading Scientific Schools,
and by the Science Support Foundation, Russia.

%%%%%%%%%%%%%%%%%%%%%%%%%%%%%%%%%%%%%%%%%%%%%%%%%%%%%%%%%%%%%%%%%%

\end{document}